\documentclass[a4paper,fleqn]{cas-sc}

\usepackage[authoryear,longnamesfirst]{natbib}
\usepackage{makecell}
\usepackage{threeparttable}
\usepackage{placeins}
\usepackage{afterpage}
\usepackage{caption}

\def\tsc#1{\csdef{#1}{\textsc{\lowercase{#1}}\xspace}}
\tsc{WGM}
\tsc{QE}

\begin{document}
\let\WriteBookmarks\relax
\def\floatpagepagefraction{1}
\def\textpagefraction{.001}

\shorttitle{}    

\shortauthors{}  

\title [mode = title]{MRBench: Evaluating Robustness and Generalization of MRI Reconstruction Foundation Models in Cardiovascular MR}  



%

\author[1]{Siying Xu}[orcid=0009-0000-1713-4951]

\cormark[1]


\ead{siying.xu@med.uni-tuebingen.de}


\credit{Conceptualization, Methodology, Software, Validation, Formal analysis, Investigation, Data curation, Visualization, Writing - original draft, Writing – review \& editing, Project administration}

\affiliation[1]{organization={Medical Image and Data Analysis (MIDAS.lab), Department of Diagnostic and Interventional Radiology, University of Tuebingen},
            city={Tuebingen},
            country={Germany}}

\author[1, 2]{Sergios Gatidis}[orcid=0000-0002-6928-4967]


\ead{sergios.gatidis@med.uni-tuebingen.de}


\credit{Supervision, Writing - review \& editing}

\affiliation[2]{organization={Department of Radiology, Stanford University},
            city={Stanford},
            state={California},
            country={USA}}

\author[1]{Thomas Küstner}[orcid=0000-0002-0353-4898]
\ead{thomas.kuestner@med.uni-tuebingen.de}
\credit{Conceptualization, Supervision, Writing - review \& editing, Funding acquisition}

\nonumnote{\textit{Abbreviations:}
ACS, autocalibration signal;
BART, Berkeley Advanced Reconstruction Toolbox;
bSSFP, balanced steady-state free precession;
CMR, cardiovascular magnetic resonance;
CS, compressed sensing;
CT, computed tomography;
ECG, electrocardiogram;
FLAIR, fluid-attenuated inversion recovery;
iid, independent and identically distributed;
I/O, input/output;
LGE, late gadolinium enhancement;
MR, magnetic resonance;
MRI, magnetic resonance imaging;
NRMSE, normalized root mean squared error;
OOD, out-of-domain;
PD, proton density;
PI, parallel imaging;
PSNR, peak signal-to-noise ratio;
RAM, Reconstruct Anything Model;
SDUM, Scalable Deep Unrolled Model;
SNR, signal-to-noise ratio;
SSIM, structural similarity index measure;
TE, echo time; 
TR, repetition time;
TSE, turbo spin echo;
VCC, virtual coil combination;
VISTA, variable density incoherent spatiotemporal acquisition.}

\cortext[1]{Corresponding author}



\begin{abstract} 
\textbf{Background:} Foundation models are emerging for magnetic resonance imaging (MRI) reconstruction, aiming to support diverse acquisition settings and imaging domains within a single model. However, differences in training data, preprocessing, and evaluation protocols make it difficult to assess and compare their generalization and operating limits directly. We aim to develop a standardized and reproducible framework for systematically evaluating pretrained MRI reconstruction foundation models.
\textbf{Methods:} We developed MRBench, a standardized, reproducible, and extensible framework for evaluating MRI reconstruction models under controlled test conditions. MRBench currently integrates three foundation models (two of which are designed for cardiovascular magnetic resonance) and supports evaluation of reconstruction performance across acquisition conditions, signal fidelity, and data domains. The primary evaluation was conducted on an in-house cardiac cine MRI dataset including 65 patients, with additional experiments using public CMRxRecon cardiac and fastMRI brain and knee datasets. Reconstruction performance was assessed quantitatively and qualitatively, and the results were made available through an interactive dashboard for comparison across models and test conditions.
\textbf{Results:} The evaluated foundation models showed distinct performance and robustness profiles across test conditions. Some changes, such as increased acceleration and measurement noise, led to gradual degradation in reconstruction quality, whereas other variations (e.g., fully-sampled center region) exposed model-specific inference failures when key acquisition or preprocessing requirements were not satisfied. Models with higher reconstruction accuracy under a given condition were not necessarily more stable as the test conditions changed, indicating that accuracy and robustness capture complementary aspects of model performance. In addition, current foundation models still showed limited generalization. Motion artifacts introduced during acquisition were difficult to remove, and reconstruction performance decreased substantially when models were applied outside their development domains, particularly to different anatomies.
\textbf{Conclusion:} MRBench enables transparent comparison of reconstruction foundation models and reveals that current foundation models still have important limitations in robustness and generalization. By systematically identifying model-specific strengths, weaknesses, and failure modes, MRBench provides evidence to guide the development of more reliable and generalizable reconstruction foundation models for clinical translation.
\end{abstract}


\begin{keywords}
MRI reconstruction \sep Foundation models \sep Cardiac cine MRI \sep Generalization \sep Benchmarking
\end{keywords}

\maketitle

\section{Introduction}\label{}

Cardiovascular magnetic resonance (CMR) provides a comprehensive assessment of cardiac anatomy and function without exposure to ionizing radiation. However, CMR examinations can be time-consuming since data are often acquired across multiple sequences, cardiac phases, imaging planes, and breath-holds. The relatively long acquisition times can compromise patient comfort, increase susceptibility to motion artifacts, and limit scanner throughput. To reduce acquisition time, accelerated MRI routinely undersamples k-space data and reconstructs images from these incomplete measurements.

Parallel imaging (PI)~\cite{pruessmann1999sense, griswold2002generalized, lustig2010spirit, uecker2014espirit} and compressed sensing (CS)~\cite{donoho2006compressed, lustig2007sparse, lustig2008compressed} are widely used reconstruction methods. PI leverages spatial encoding information from multiple receiver coils to recover missing information, while CS exploits prior knowledge of image structure, such as sparsity, through regularized reconstruction. Despite their widespread use, PI and CS remain limited by low acceleration factors, long iterative optimization processes, and manual parameter tuning.

Deep learning has substantially improved accelerated MRI reconstruction by learning priors directly from data, enabling faster reconstruction and improved image quality. However, most reconstruction networks are developed for specific anatomies, acquisition protocols, or sampling conditions, which limits their performance when applied outside the training distribution. Foundation models have recently been explored as a way to extend reconstruction models across diverse imaging scenarios. These models are typically pretrained on large-scale and diverse datasets to support a broader range of MRI reconstruction applications within a single model~\cite{zhang2024foundation, lyu2025crunet, tavakoli2025vtransformer, he2025omnimri, wang2025enabling, terris2025reconstruct, wang2026sdum}. Some are developed within a specific imaging domain, such as CMR, and designed to support multiple sequences, sampling patterns, and acquisition settings, while others aim to solve a broader range of inverse problems.

Despite these advances, it remains unclear when these models can be reliably applied and how well they perform when acquisition conditions or data characteristics differ from their development settings~\cite{knoll2019assessment, darestani2022test, antun2020instabilities}. Differences in training data, preprocessing and inference pipelines, and evaluation protocols further make their reported performance difficult to compare directly~\cite{shimron2022implicit}. A systematic comparative evaluation is therefore needed to characterize their generalizability, robustness, and potential failure modes.

To address these issues, we introduce MRBench, a standardized and reproducible framework for systematically evaluating pretrained MRI reconstruction foundation models under controlled test conditions. We aim to establish an extensible framework that can continuously incorporate newly developed foundation models and datasets as the field evolves. Currently, MRBench integrates three recently proposed reconstruction foundation models and investigates their performance across variations in acquisition conditions, signal fidelity, and data domains. The evaluation focuses on model robustness and the identification of potential failure modes. In this work, we focus on the analysis of CMR data. MRBench is therefore designed to support transparent and extensible benchmarking, providing evidence for model selection and the development of more generalizable MRI reconstruction models. 

\section{Method}
\subsection{MRBench framework}  
MRBench is a standardized and reproducible framework for systematically evaluating foundation models for MRI reconstruction. It brings MRI data, test conditions, foundation models, and performance evaluation into a single benchmarking framework. The framework is designed to support diverse MRI datasets and foundation models while providing a consistent basis for studying how model performance changes across acquisition, signal, and data domains.

\begin{figure}
	\centering
	\includegraphics[width=\textwidth]{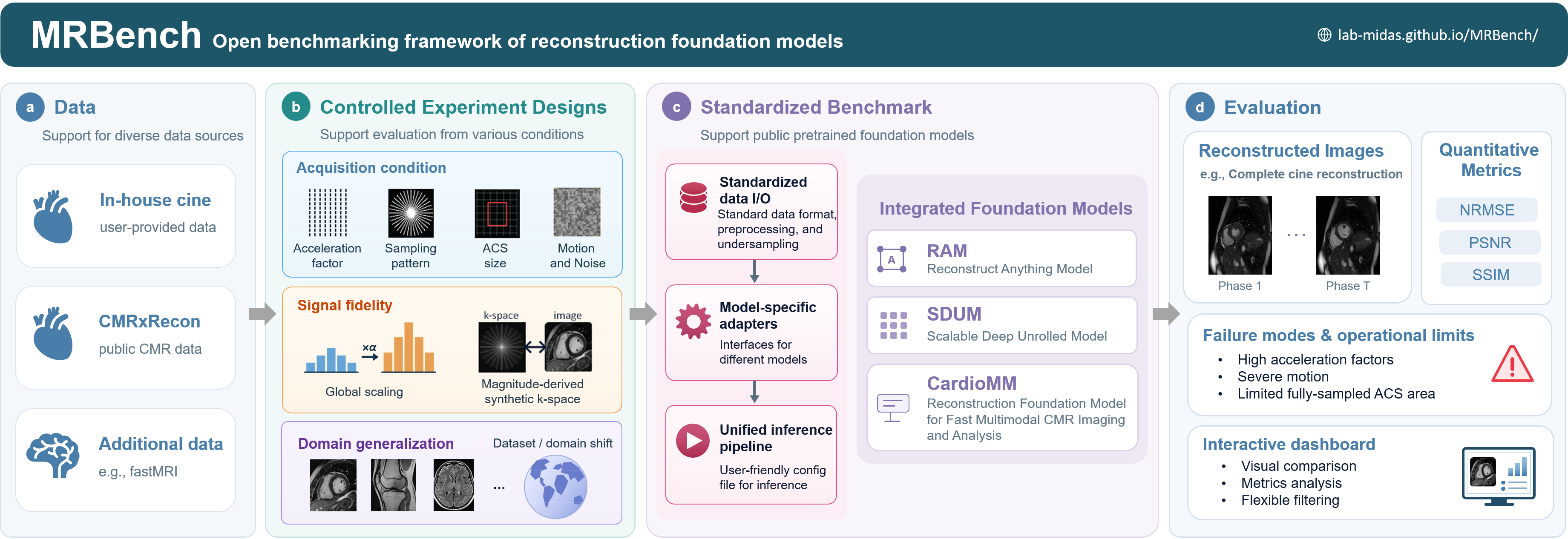}
	\caption{\textbf{Overview of the MRBench framework.} (a) Support for diverse MRI datasets, with a primary focus in this work on cardiac cine MRI. (b) Controlled experimental designs for assessing acquisition conditions, signal fidelity, and domain generalization. (c) Standardized benchmark that investigates three pretrained MRI reconstruction models (RAM, SDUM, and CardioMM) through model-specific adapters, standardized data interfaces, and a unified inference pipeline. (d) Qualitative and quantitative evaluation (NRMSE, PSNR, and SSIM), with an interactive dashboard for visualizing reconstruction performance and inference failures across experimental conditions.}
	\label{FIG:framework}
\end{figure}

As illustrated in Fig~\ref{FIG:framework}, MRBench connects four elements: input data, controlled experiments, model inference, and evaluation. MRBench supports MRI data from different sources and with different dimensions. In this work, the framework primarily investigates multi-coil cardiac cine MRI. The controlled experiments introduce representative variations that may affect reconstruction in practice. These variations are grouped into three categories: acquisition conditions, signal fidelity, and data domain generalization, which will be introduced in detail in Section~\ref{sec:method_exp}. 

All reconstructions are performed using pretrained foundation models without any additional training or fine-tuning. This setup reflects a practical clinical workflow in which existing foundation models are directly applied to acquired data. MRBench currently includes RAM~\cite{terris2025reconstruct}, SDUM~\cite{wang2026sdum}, and CardioMM~\cite{wang2025enabling}. Each model is integrated through a model-specific adapter that converts the MRBench data format into the inputs required by the inference pipeline. Reconstruction outputs are subsequently evaluated using qualitative and quantitative measures. These results are used to characterize the operating boundaries and potential failure modes of each foundation model, providing insight into its transferability to routine clinical practice. To facilitate comparison across models and experimental conditions, MRBench provides an interactive dashboard for visualizing and exploring the benchmark results.

\subsection{Evaluated foundation models} 
MRBench currently integrates three foundation models for reconstruction: Reconstruct Anything Model (RAM)~\cite{terris2025reconstruct}, Scalable Deep Unrolled Model (SDUM)~\cite{wang2026sdum}, and CardioMM~\cite{wang2025enabling}. The three models were selected because they represent different scopes and strategies for achieving reconstruction generalization in CMR. RAM was developed for multiple inverse problems, whereas SDUM and CardioMM focus on reconstructions within CMR. Table~\ref{tbl:model_characteristic} summarizes the key characteristics of each model.

RAM is a general reconstruction model for solving diverse inverse problems, including MRI, computed tomography (CT), deblurring, inpainting, and super-resolution~\cite{terris2025reconstruct}. Its MRI training data were derived from the multi-coil fastMRI brain dataset~\cite{zbontar2018fastmri}, which were converted into single-coil images using virtual coil combination. MRI reconstruction with $4\times$ and $8\times$ Cartesian undersampling masks was included in the training. For inference, RAM requires undersampled k-space together with the corresponding forward operator. For multi-coil MRI, this operator is defined by the undersampling mask and coil sensitivity maps.

SDUM was developed for universal cardiac MRI reconstruction~\cite{wang2026sdum}. Its training data includes CMRxRecon2024 and CMRxRecon2025 datasets~\cite{wang2025cmrxrecon2024}, covering a broad range of CMR applications (e.g., cine, T1/T2 mapping, LGE, perfusion, and flow imaging), sampling patterns, and acceleration factors. Note that a brain-specific SDUM was additionally trained on fastMRI brain data, which was not included for this analysis in MRBench. Three released model sizes ("Small", "Base", and "Large") are supported in MRBench and can be selected directly for inference through the model configuration. SDUM estimates coil sensitivity maps internally and therefore does not require them to be provided by the user. In addition to the undersampled k-space and sampling mask, inference requires acquisition information describing the sampling pattern, nominal acceleration factor, and acquisition type (e.g., "uniform", "8", "cine").

CardioMM~\cite{wang2025enabling} is a reconstruction foundation model for fast multimodal CMR imaging and analysis. It was developed using MMCMR-427K, a large-scale multi-population, multi-disease, multi-center, multi-vendor, and multimodal CMR database. Similar to SDUM, CardioMM estimates coil sensitivity maps internally. In addition to the undersampled k-space and undersampling mask, CardioMM requires two text prompts for reconstruction: an undersampling prompt specifying the sampling pattern and acceleration factor (e.g., "Undersampling: $8\times$ uniform."), and a metadata prompt describing the age group, field strength, scanner vendor, and imaging modality (e.g., "Adult cardiac MRI reconstruction. Vendor: 1.5T, Siemens, Aera. Modality: cine, short-axis."). 

\begin{table}[width=.9\linewidth,cols=2,pos=h]
\caption{Demographic characteristics of the in-house cardiac cine study cohort}\label{tbl:data_characteristic}
\begin{tabular*}{\tblwidth}{@{} LL@{}}
\toprule
Characteristic & Study cohort \\
\midrule
Participants, n & 65  \\
Age, years, mean ± SD (range) & 45.9 ± 17.6 (17 to 83) \\
Female sex, $n$ (\%) & 24 (36.9\%) \\
Male sex, $n$ (\%)       & 41 (63.1\%) \\
\bottomrule
\end{tabular*}
\end{table}

\begin{table}[width=.9\linewidth,cols=4,pos=h]
\begin{threeparttable}
\caption{Development domains, acquisition exposure, and inference assumptions of the evaluated reconstruction foundation models.}\label{tbl:model_characteristic}
\begin{tabular*}{\tblwidth}{@{}cccc@{}}
\toprule
Characteristic & RAM\tnote{a} & SDUM\tnote{a} & CardioMM\tnote{a}  \\ 
\midrule
Model scope & 
\makecell{General reconstruction model\\ for computational imaging} & 
\makecell{Universal CMR\\ reconstruction model} & 
\makecell{Generalist multimodal\\ CMR reconstruction model} \\
\midrule
Training data\tnote{b} &
\begin{tabular}{@{}c@{\hspace{3pt}}c@{}}
    MRI: & fastMRI brain \\
    CT:  & LIDC-IDRI\\
    Other:  & LSDIR
\end{tabular} &
\begin{tabular}{@{}c@{\hspace{3pt}}c@{}}
    \multirow{3}{*}{MRI:} & CMRxRecon 2024 \\
                          & CMRxRecon 2025 \\
                          & fastMRI\tnote{c}
\end{tabular} &
\begin{tabular}{@{}l@{\hspace{3pt}}l@{}}
    MRI: & MMCMR-427K
\end{tabular} \\
\midrule
Training anatomy & Brain & Cardiac & Cardiac \\
\midrule
Training coil setting & Single-coil\tnote{d} & Multi-coil & Multi-coil \\
\midrule
Sampling exposure & Cartesian\tnote{e} & 
\makecell{Uniform, Gaussian,\\ Pseudo-radial} & 
\makecell{Uniform, Random,\\ Pseudo-radial} \\
\midrule
Acceleration exposure & $4\times$, $8\times$ & $8\times$, $16\times$, $24\times$ & $4\times$, $8\times$, $16\times$, $24\times$ \\
\midrule
Required inputs &
\makecell{undersampled k-space, \\ undersampling mask, \\ coil sensitivity maps} &
\makecell{undersampled k-space, \\ undersampling mask, \\ sampling information}  &
\makecell{undersampled k-space, \\ undersampling mask, \\ sampling information, \\ scan metadata} \\
\midrule
Sensitivity map & External & Internal & Internal \\
\midrule
ACS requirement & No & Yes & Yes \\
\bottomrule
\end{tabular*}
\begin{tablenotes}[flushleft]
\footnotesize
\item[a] References for the evaluated foundation models: RAM~\cite{terris2025reconstruct}, SDUM~\cite{wang2026sdum}, CardioMM~\cite{wang2025enabling}.
\item[b] References for the training datasets are as follows: fastMRI~\cite{zbontar2018fastmri}, LIDC-IDRI~\cite{armato2011lung}, LSDIR~\cite{li2023lsdir}, CMRxRecon 2024~\cite{wang2025cmrxrecon2024}, CMRxRecon 2025~\cite{CMRxRecon2025}, MMCMR-427K~\cite{wang2025enabling}.
\item[c] FastMRI was used only to train a separate SDUM model to evaluate generalizability to other anatomical regions.
\item[d] RAM used virtual coil-combined images derived from multi-coil fastMRI data~\cite{terris2025reconstruct}.
\item[e] The specific Cartesian sampling pattern used for RAM training was not reported in the paper or released code.
\end{tablenotes}
\end{threeparttable}
\end{table}

\subsection{Controlled experiment designs}\label{sec:method_exp} 
This section describes the controlled experiments supported in MRBench together with the investigated parameter ranges in this work. The experiments are designed to simulate variations and perturbations that may arise in scientific and clinical practice and are organized into three groups: acquisition conditions, signal fidelity, and domain generalization.

\subsubsection{Acquisition conditions}
\textbf{Acceleration factor.} MRBench supports testing with flexible acceleration factors. To evaluate model robustness to increasing undersampling severity, we evaluated acceleration factors of $R=8/16/24$. The specified $R$ represents the nominal acceleration. For sampling patterns that include fully-sampled autocalibration signal (ACS) data, the resulting net acceleration factor $R_{net}$ can be lower because of the additional acquired ACS samples. For models requiring acceleration information in text, the acceleration factor refers to the nominal acceleration.

\textbf{Sampling pattern.} MRBench can generate different sampling masks, including 1D/2D masks for static imaging and k-t sampling masks for dynamic imaging. It also supports user-provided masks. In this study, we mainly evaluated three dynamic sampling patterns: kt-uniform, kt-Gaussian, and kt-radial. The kt-radial pattern is pseudo-radial sampling, in which radial spokes are rasterized onto a Cartesian k-space grid. Since SDUM and CardioMM require the nominal acceleration factor $R$ as an input, we generated the sampling masks using the same nominal $R$ across different sampling patterns. As described above, differences in ACS coverage can result in different net acceleration factors $R_{\mathrm{net}}$.

\textbf{ACS dependence.} MRBench allows the size of the fully-sampled ACS to be specified independently of the sampling pattern, and also supports sampling masks without a fully-sampled ACS region, such as VISTA~\cite{ahmad2015variable}. Since SDUM and CardioMM rely on ACS data for internal coil sensitivity estimation, we progressively reduced the ACS size to evaluate this dependence: $n_{\mathrm{calib}}=20,15,10,5,1$ center lines. We used kt-uniform Cartesian sampling and adjusted the spacing between k-space lines for each ACS size to keep a similar net acceleration factor, $R_{net}\approx6$. In addition, we tested model compatibility with VISTA sampling, and repeated the reconstruction after adding a fully-sampled ACS region ($n_{\mathrm{calib}}=20$ lines) to the same VISTA mask for comparison.

\textbf{Noise.}
During acquisition, the measured k-space signal is corrupted by thermal noise, mainly from random fluctuations of charges in the patient's conductive tissue and, to a lesser extent, from the receiver coils and preamplifiers. This noise is well modeled as additive, zero-mean, white Gaussian noise in the real and imaginary channels of the retrospectively undersampled multi-coil k-space~\cite{ekanayake2025cl, virtue2017empirical}. To evaluate model robustness to measurement noise, we defined the noise standard deviation $\sigma_c$ for an undersampled measurement $y_0$ and a target SNR $s$ [dB] as follows:
\begin{equation}
    \sigma_c =\sqrt{\frac{1}{N}\sum_{i\in\Omega}|y_{0,i}|^2} 10^{-s/20},
\end{equation}
where $\Omega$ denotes the set of acquired k-space locations across all dimensions, and $N=\left | \Omega \right |$ is the total number of acquired samples. Independent complex Gaussian noise is then generated as:
\begin{equation}
    n_i=\frac{\sigma_c}{\sqrt{2}}(u_i+\mathrm{j}v_i),\qquad u_i,v_i \overset{\mathrm{iid}}{\sim}\mathcal{N}(0,1),
\end{equation}
where $i$ is an entry in $\Omega$, $\mathrm{j}$ denotes the imaginary unit, \(u_i\) and \(v_i\) are independent standard normal random variables generating the real and imaginary noise components, and $\mathbb{E}|n_i|^2=\sigma_c^2$. Noise was added only to the sampled k-space locations. We evaluated target measurement SNRs of 35, 30, and 25 dB, with lower SNR corresponding to higher measurement noise during acquisition.

\textbf{Motion.} In gated, segmented acquisitions, residual respiratory motion from incomplete breath-holds, heart-rate variability and bulk motion cause inconsistencies between k-space segments of the same cardiac phase. The binned data therefore violate the quasi-periodicity assumption, which leads to residual aliasing and temporal blurring. In this work, we simulated two representative motion sources: cardiac mistriggering and respiratory motion. 

To simulate cardiac phase inconsistencies, we model electrocardiogram (ECG) mistriggering, in which k-space data acquired at one cardiac phase are assigned to a different phase because of erroneous trigger detection. We hence randomly selected a fraction $f$ of the sampled phase-encoding lines in each slice and cardiac phase, and replaced them with the corresponding $k_y$ lines from a different randomly selected cardiac phase. We evaluated fractions $f=0,0.1,0.2,0.3$, with $f=0$ representing no added motion.

Cardiac cine MRI is commonly acquired as a stack of 2D slices, with each slice (or a small group of slices) acquired in a separate breath-hold to minimize the impact of respiratory motion. However, imperfect breath-holding or diaphragmatic drift can still displace the heart within a breath-hold, potentially causing k-space lines to be collected at different respiratory positions~\cite{larson2005preliminary}. In our cardiac cine acquisition, $x$ is the frequency-encoding direction, oriented superior-inferior, whereas $y$ denotes the phase-encoding direction, oriented anterior-posterior. Without loss of generality, we modeled respiratory motion as an in-plane translation along the $x$ direction, with the displacement varying sinusoidally across phase-encoding $k_y$ lines. The displacement assigned to the $i$-th $k_y$ line was:
\begin{equation}
    \Delta_x(i)=A\sin\left(2\pi q\frac{i}{N_y}\right),
\end{equation}
where $i$ denotes the index of $k_y$ lines, $N_y$ is the total number of $k_y$ lines, $A$ is the maximum displacement in pixels, and $q=4$ specifies the number of sinusoidal cycles across the full $k_y$ grid. The translation was implemented in k-space through the corresponding phase modulation along $k_x$ and for various phase-encodings $k_y$:
\begin{equation}
    \tilde{K} _{z,t}(\nu_x,i,c)=K_{z,t}(\nu_x,i,c)\exp\left[-2\pi\mathrm{j}\nu_x\Delta_x(i)\right],
\end{equation}
where $z$, $t$, $c$ denote the slice, cardiac phase, and coil, respectively; $\nu_x$ denotes the centered spatial frequency along $k_x$, expressed in cycles per pixel. We evaluated displacement amplitudes of $A=0,2,5,10$ pixels, with $A=0$ corresponding to no added respiratory motion.

\subsubsection{Signal fidelity}
\textbf{Global scaling.} The numerical scale of MRI measurements can vary across datasets and preprocessing pipelines. To evaluate the sensitivity to the input scale range, we scaled the input k-space data globally by a constant factor $\alpha$:
\begin{equation}
    \widetilde{y}=\alpha y,
\end{equation}
where $y$ is the original k-space and $\alpha$ denotes the global scaling factor. Five scaling factors were tested: $\alpha=0.01,0.1,1,10,100$, with $\alpha=1$ representing the original signal.

\textbf{Magnitude-derived synthetic k-space.} MRI datasets are sometimes only available as magnitude images, from which synthetic k-space under a zero-phase assumption may be generated. Hence, the resulting non-zero phase in k-space does not reflect the true k-space phase information. To evaluate model sensitivity to this difference, we generated synthetic k-space from the magnitude of the reference image.

\subsubsection{Domain generalization}
In- and out-of-domain (OOD) behavior was evaluated across data sources, cardiac views, and imaging anatomies. In this study, in-domain refers to imaging characteristics consistent with the model's training data, while OOD refers to imaging characteristics not encountered during training, such as unseen anatomies or data from a different center. We tested all three models on in-house short-axis cardiac cine MRI, CMRxRecon 2024 cine data including short-axis and long-axis (2-chamber, 3-chamber, and 4-chamber) views~\cite{wang2025cmrxrecon2024}, and multi-coil fastMRI brain and knee data~\cite{zbontar2018fastmri}. These datasets represented different degrees of domain shift for each model. For RAM, cardiac cine and fastMRI knee data represented unseen anatomies, while multi-coil fastMRI brain shared the training anatomy but differed from the single-coil data used during training. For SDUM and CardioMM, brain and knee were out-of-domain anatomies, whereas the in-house cine data belonged to the familiar cardiac imaging domain but were acquired from a different center.

\subsection{Datasets} 
For evaluation, we used a combination of in-house data and publicly available datasets: CMRxRecon~\cite{wang2025cmrxrecon2024} and fastMRI~\cite{zbontar2018fastmri}. The in-house 2D cardiac cine MRI includes 65 patients (demographic characteristics in Table~\ref{tbl:data_characteristic}), acquired on a 1.5T scanner (MAGNETOM Aera, Siemens Healthineers, Erlangen, Germany) using a balanced steady-state free precession (bSSFP) sequence. Short-axis cine images covering the left ventricle were acquired under multiple breath-holds. The imaging parameters were TE / TR = 1.06 / 2.12 ms, flip angle = $52^\circ$, bandwidth = 915 Hz/px, spatial resolution = $1.9 \times 1.9$ mm$^2$, slice thickness = 8 mm, and temporal resolution = 40 ms. Each cine series contained 25 cardiac phases. The study was approved by the Ethik-Kommission an der Medizinischen Fakultät der Eberhard-Karls-Universität und am Universitätsklinikum Tübingen (approval no. 426/2021BO1), and written informed consent was obtained from all participants.

Coil sensitivity maps were estimated from 24 central k-space calibration lines using ESPIRiT~\cite{uecker2014espirit} and compressed to 15 channels using the Berkeley Advanced Reconstruction Toolbox (BART)~\cite{uecker2015berkeley}. These sensitivity maps were provided only to reconstruction models that require external coil sensitivity information, such as RAM.


The selected CMRxRecon 2024 data were acquired in short-axis and long-axis using a bSSFP sequence on a 3T Siemens MAGNETOM Vida scanner. The fastMRI knee example shown in the results was acquired using a coronal proton density-weighted turbo spin echo (TSE) sequence with fat suppression on a 3T Siemens MAGNETOM Prisma fit scanner, and the brain example was acquired using an axial fluid-attenuated inversion recovery (FLAIR) turbo spin-echo sequence on a 3T Siemens MAGNETOM Skyra scanner. For these publicly available datasets, the coil sensitivity maps required by RAM were estimated from 20 central k-space calibration lines using ESPIRiT~\cite{uecker2014espirit}. This sensitivity map estimation method is also available as a configurable option offered by MRBench. 

\subsection{Evaluation metrics} 
Reconstruction quality was evaluated using normalized root mean squared error (NRMSE), peak signal-to-noise ratio (PSNR), and structural similarity index measure (SSIM). NRMSE was normalized by the $L_2$ norm of the reference image, while PSNR and SSIM used a fixed data range of 1. SSIM was calculated using a $7\times7$ window with uniform weights. Before metric calculation, the reconstructed and reference cine data were converted to magnitude images and independently normalized by their respective 99.5th percentile intensities without clipping values above 1. Metrics were calculated for each 2D frame and were equally averaged to obtain one value per subject. Cohort-level distributions were summarized using these subject-level metrics.

\subsection{Implementation details} 
All three models were evaluated in MRBench using pretrained weights without task-specific fine-tuning. All RAM results reported in this study were obtained using the RAM implementation in DeepInverse 0.3.5~\cite{tachella2025deepinverse}. Since RAM is designed for 2D images, each slice and cardiac phase of the cine MRI was reconstructed independently. SDUM provides three model variants of increasing capacity: "Small", "Base", and "Large"~\cite{wang2026sdum}. We used the "Base" variant for all results reported in this study to balance reconstruction performance and computational efficiency. All three variants are incorporated into MRBench and can be selected through the model configuration. Similar to RAM, CardioMM also performs 2D reconstruction for each slice and cardiac phase independently. For SDUM and CardioMM, the required sampling and acquisition information was provided as additional inputs following their original implementations.

The MRBench repository and interactive dashboard are available at \url{https://github.com/lab-midas/MRBench} and \url{https://lab-midas.github.io/MRBench/}, respectively. Additional implementation details and results are available in the public dashboard.

\section{Results}
\subsection{Acceleration robustness}
\begin{figure}
	\centering
	\includegraphics[width=\textwidth]{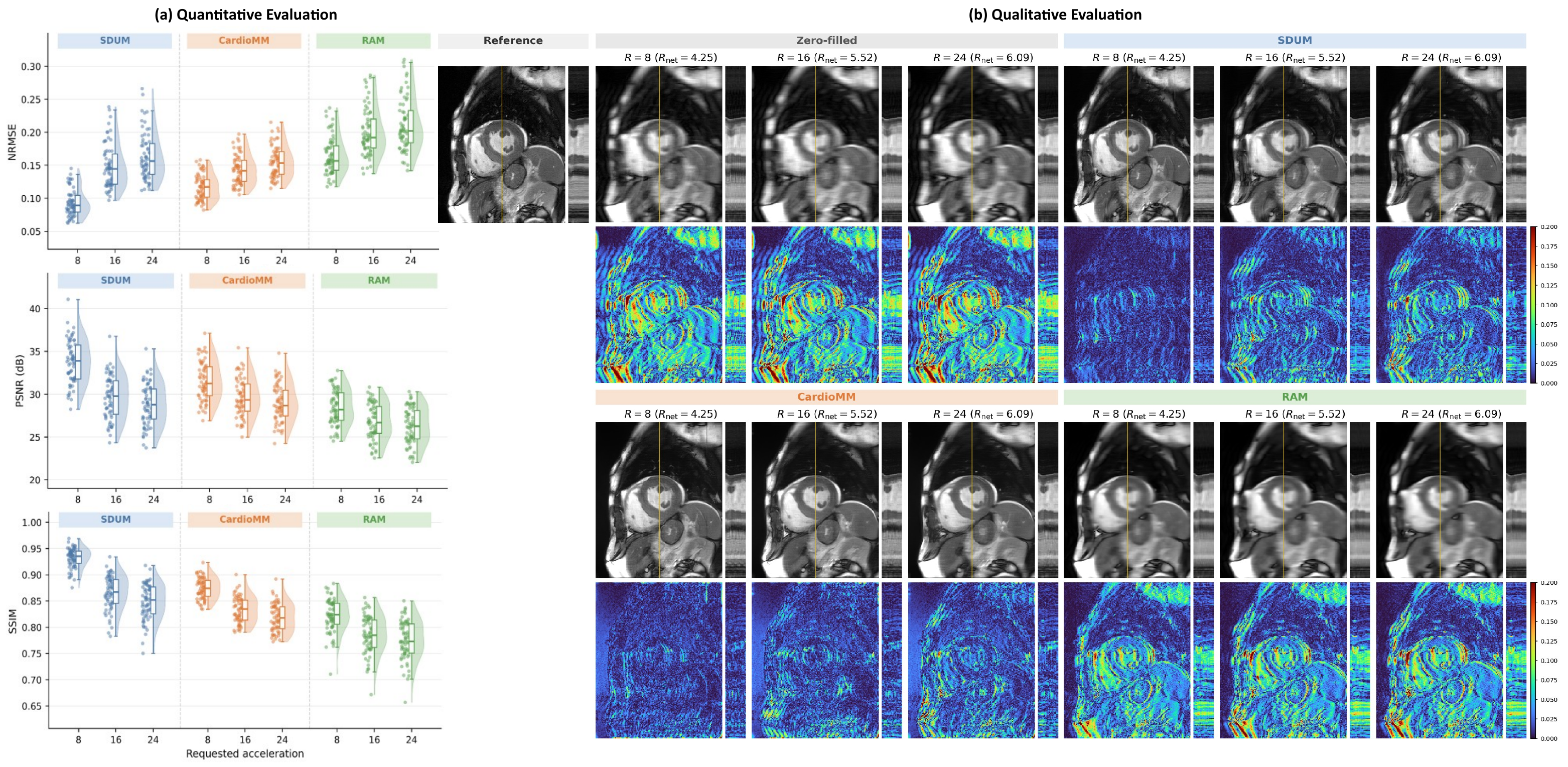}
	\caption{\textbf{Reconstruction performance across different acceleration factors.} (a) Raincloud plots showing NRMSE, PSNR, and SSIM for SDUM, CardioMM, and RAM across 65 patients from an in-house cardiac cine MRI dataset using kt-uniform undersampling at nominal acceleration factors of $R=8,16,$ and $24$. The plots combine individual data points, distribution densities, and boxplots. (b) Representative results from one patient, including the fully-sampled reference, zero-filled reconstructions, and reconstructed images from the three models, along with their corresponding absolute error maps. The yellow lines in the x-y images indicate the corresponding y-t planes. Effective net acceleration factors ($R_{net}$) are indicated in parentheses above each reconstruction.}
	\label{FIG:result_acc}
\end{figure}

As shown in Fig.~\ref{FIG:result_acc}, we evaluated reconstruction performance using kt-uniform undersampling at nominal acceleration factors of $R=8,16,24$. The same sampling mask was applied to all three models. We observed that all three models showed progressively decreased reconstruction quality as acceleration increased. SDUM showed the most pronounced change between $R=8$ and $R=16$, with mean SSIM decreasing from 0.932 to 0.861. At $R=8$, SDUM showed the best performance, with the lowest NRMSE and the highest PSNR and SSIM, but this advantage narrowed as acceleration increased. At $R=24$, SDUM retained a higher SSIM compared to CardioMM (0.848 vs. 0.818), but had a slightly higher NRMSE (0.163 vs. 0.155) and lower PSNR (28.8 vs. 29.0 dB).

Qualitatively, Fig.~\ref{FIG:result_acc}(b) shows representative reconstructions from one patient with clinically diagnosed dilated cardiomyopathy and a dilated left ventricle. RAM exhibited overall blurring and loss of anatomical detail, whereas SDUM and CardioMM better preserved overall anatomy. However, residual artifacts became more apparent as acceleration increased. In the SDUM reconstruction at $R=16$, concentric residual artifacts were visible around the myocardial boundary. Fine anatomical structures such as papillary muscles also became blurrier in both SDUM and CardioMM reconstructions.

\subsection{Sampling pattern sensitivity}
\begin{figure}
	\centering
	\includegraphics[width=\textwidth]{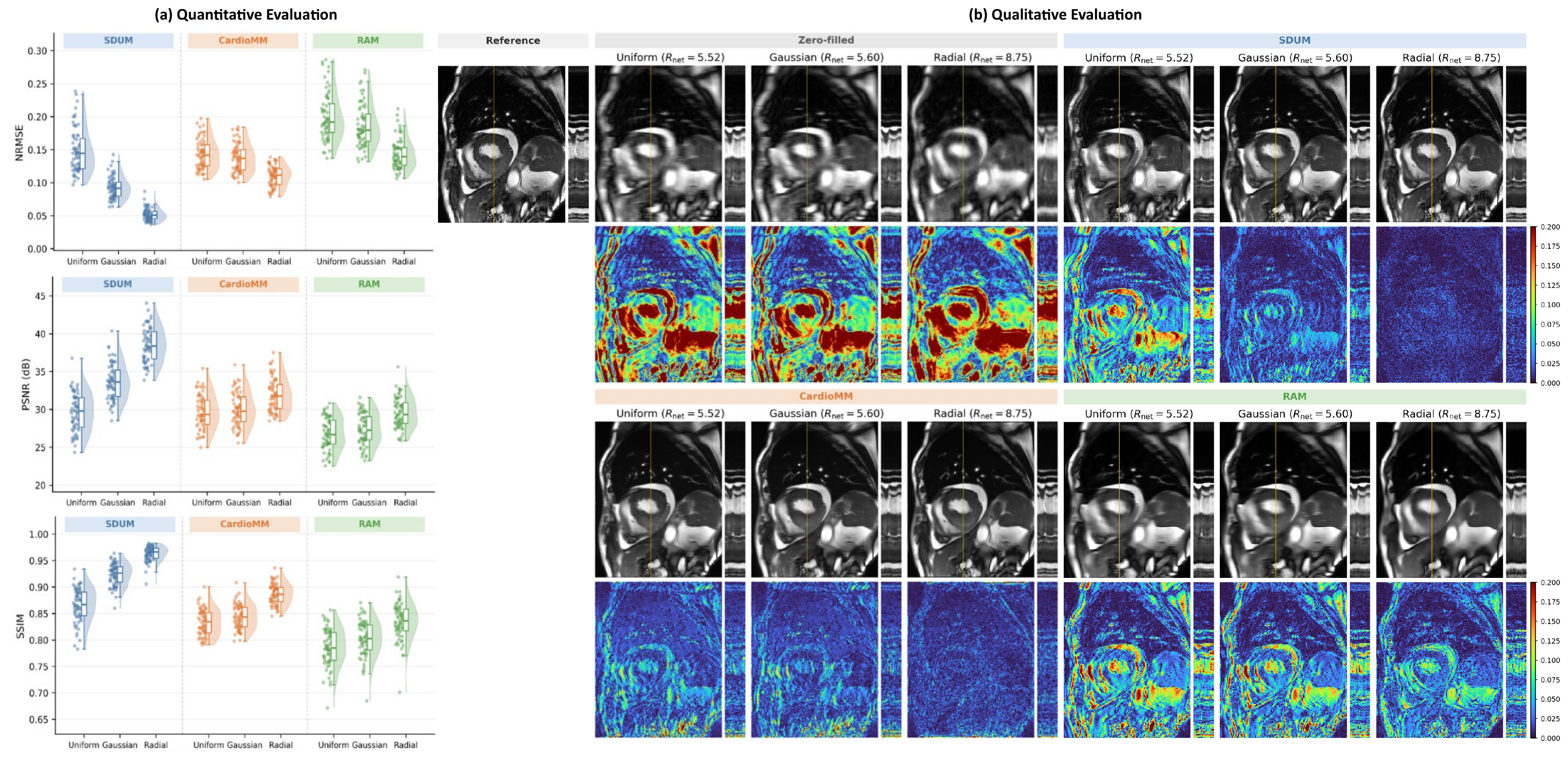}
	\caption{\textbf{Reconstruction performance across different sampling patterns.} (a) Raincloud plots showing NRMSE, PSNR, and SSIM for SDUM, CardioMM, and RAM across 65 patients from an in-house cardiac cine MRI dataset, using kt-uniform, kt-Gaussian, and pseudo kt-radial undersampling for an investigated nominal acceleration factor of R=16. The plots combine individual data points, distribution densities, and boxplots. (b) Representative results from one patient, including the fully-sampled reference, zero-filled reconstructions, and reconstructed images from the three models, along with their corresponding absolute error maps. The yellow lines in the x-y images indicate the corresponding y-t planes. Effective net acceleration factors ($R_{net}$) are indicated in parentheses above each sampling pattern.}
	\label{FIG:result_sampling}
\end{figure}

Fig.~\ref{FIG:result_sampling} shows reconstruction results using kt-uniform, kt-Gaussian, and kt-radial undersampling, all with the same nominal acceleration factor of $R=16$. Across all models, kt-radial yielded the best quantitative results, despite having the highest effective acceleration $R_{net}$ among the three patterns. SDUM was the most sensitive to changes in sampling pattern, with mean SSIM increasing from 0.866 with kt-uniform sampling to 0.924 with kt-Gaussian sampling and 0.963 with kt-radial sampling. CardioMM and RAM showed smaller differences between kt-uniform and kt-Gaussian sampling, while both still showed clear improvements with kt-radial sampling.

The qualitative evaluation in Fig.~\ref{FIG:result_sampling}(b) presented the reconstructions of a patient with a clinically reported intramyocardial mass suspicious for malignancy. The results showed the same sampling-dependent differences. With SDUM and CardioMM, kt-radial sampling enabled better reconstructions with reduced residual artifacts. RAM also improved with kt-radial sampling, although its reconstructions remained visibly blurred.

\subsection{ACS dependence}
\begin{figure}
	\centering
	\includegraphics[width=\textwidth]{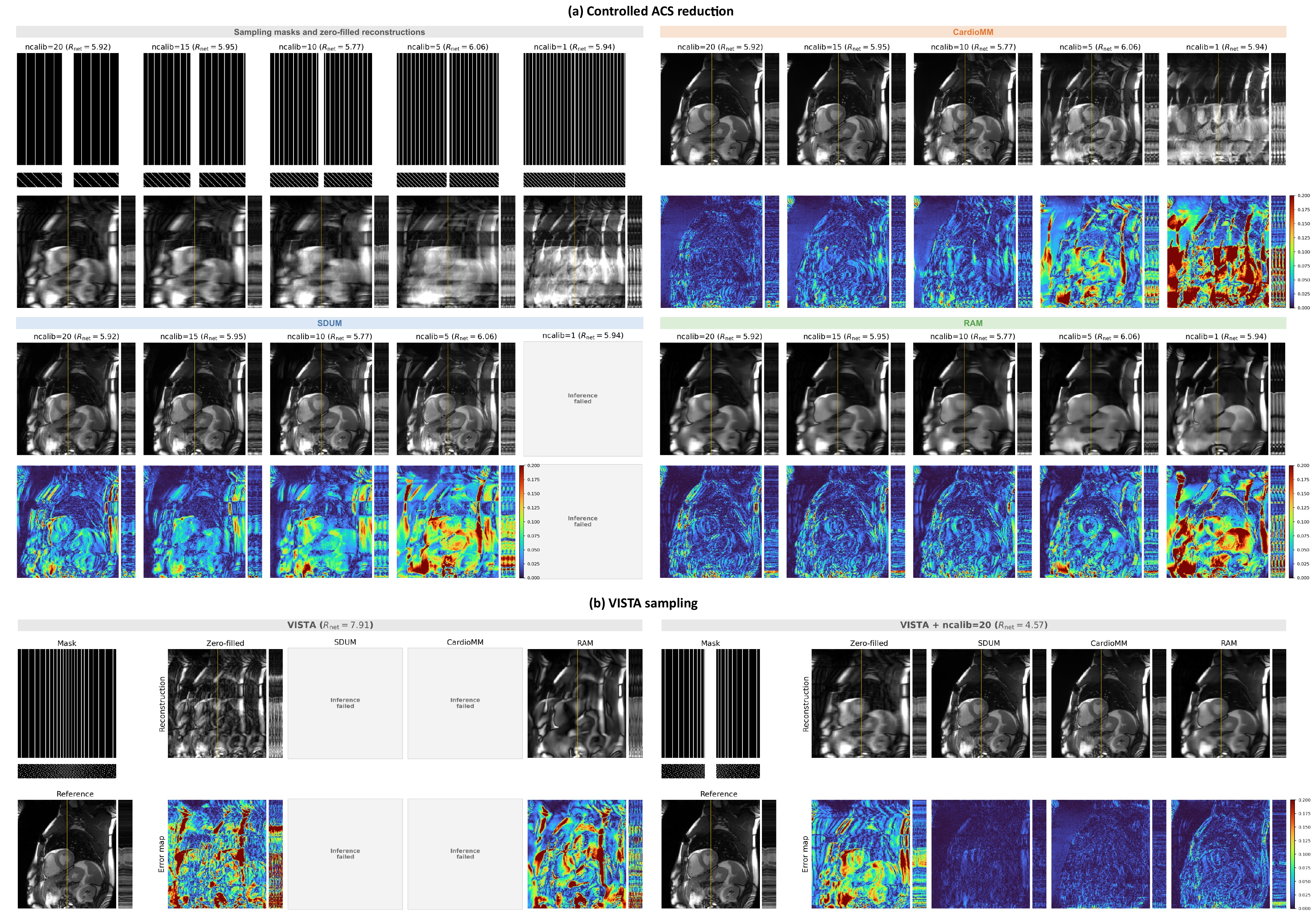}
	\caption{\textbf{Reconstruction performance under different ACS conditions.} (a) Controlled reduction of the fully-sampled autocalibration signal (ACS) region using kt-uniform undersampling. ACS sizes of $n_{\mathrm{calib}}=20,15,10,5,$ and $1$ lines were evaluated ($R_{\mathrm{net}}\approx6$). For each ACS size, the corresponding sampling mask and zero-filled reconstruction are shown together with the CardioMM, SDUM, and RAM reconstructions and their respective absolute error maps. (b) Reconstruction using VISTA sampling ($R=8$) and after adding a fully-sampled ACS region with $n_{\mathrm{calib}}=20$ lines. The yellow lines in the x-y images indicate the corresponding y-t planes.}
	\label{FIG:result_acs}
\end{figure}

Fig.~\ref{FIG:result_acs} shows that reducing the fully-sampled ACS region has a clear impact on reconstruction performance. The effect is more pronounced for SDUM and CardioMM, both of which rely on the ACS region for their internal coil sensitivity estimation. At $n_{\mathrm{calib}}=20$ and $15$ lines, CardioMM maintained relatively good reconstruction quality, whereas SDUM showed more variable behavior, with visible artifacts in the right ventricle already present at $n_{\mathrm{calib}}=20$ lines. When $n_{\mathrm{calib}}$ was reduced to 10 lines, degradation became clearer, and at $n_{\mathrm{calib}}=5$ lines, both models showed substantial degradation with noticeable residual artifacts. At $n_{\mathrm{calib}}=1$ line, CardioMM could no longer recover the overall cardiac anatomy, while SDUM failed during inference. Although RAM could perform reconstructions across all tested ACS sizes, the reconstructed image quality was poor. The images became increasingly blurred as the ACS region was reduced, with progressive loss of anatomical delineation.

This dependence on ACS size was further demonstrated using VISTA sampling, which does not contain a fully-sampled calibration region for each cardiac phase or slice (Fig.~\ref{FIG:result_acs}(b)). Under VISTA sampling ($R=8$), both SDUM and CardioMM failed during inference. RAM remained executable, but the resulting reconstruction was dominated by severe residual aliasing and was of limited diagnostic use. In contrast, when a fully-sampled ACS region was added to this VISTA mask, both SDUM and CardioMM were able to reconstruct images with markedly improved reconstruction quality. An appropriate calibration region hence not only affects reconstruction quality, but is also important for reliable inference. These findings show that current foundation models are very sensitive to the existence of a fully-sampled k-space center, i.e., to their sampling mask type seen during training.

\subsection{Noise robustness}
The robustness of the three models to measurement noise is shown in Supplementary Fig.~S1.  SDUM was the most sensitive to changes in measurement SNR, with mean SSIM increasing from 0.85 at 25 dB to 0.92 at 35 dB. CardioMM showed smaller changes over the same range (mean SSIM from 0.83 to 0.87), while RAM remained comparatively stable at a mean SSIM of 0.82 but had lower overall reconstruction quality. Visual differences across the tested noise levels were subtle for all three models.

\subsection{Motion sensitivity}
\begin{figure}
    \centering
    \includegraphics[width=\textwidth]{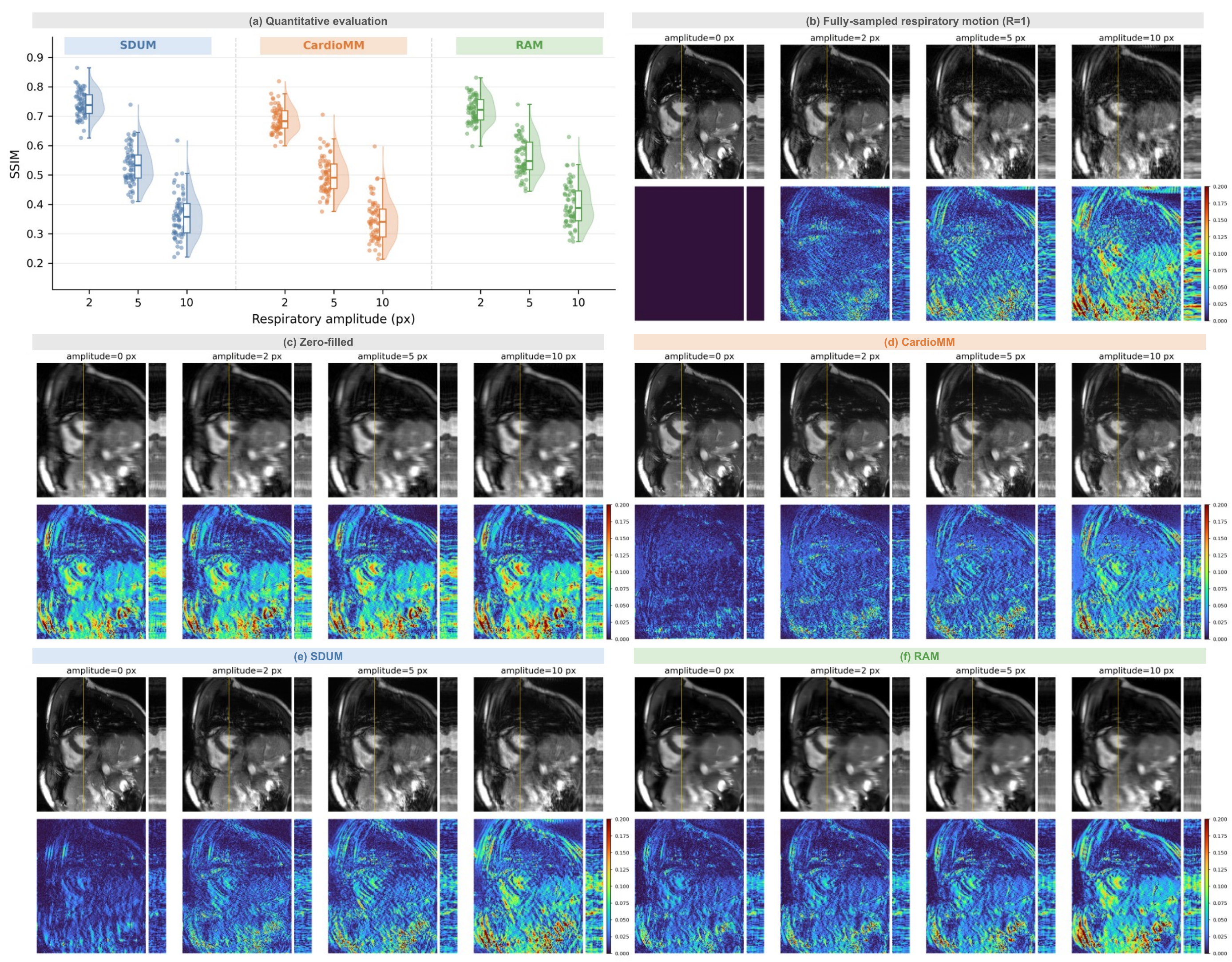}
    \caption{\textbf{Reconstruction performance under simulated respiratory motion.} (a) Raincloud plots showing SSIM distributions for SDUM, CardioMM, and RAM across 65 patients using kt-uniform undersampling at a nominal acceleration factor of $R=8$, with respiratory motion amplitudes of 2, 5, and 10 pixels. The plots combine individual data points, distribution densities, and boxplots. (b) Fully-sampled images with simulated respiratory motion. (c-f) Representative results from one patient, including zero-filled reconstructions and reconstructions from CardioMM, SDUM, and RAM, respectively, along with their corresponding absolute error maps. The respiratory motion amplitude denotes the maximum in-plane displacement, which varied sinusoidally across the phase-encoding lines. The yellow lines in the x-y images indicate the corresponding y-t planes.}
    \label{fig:results_motion_respiratory}
\end{figure}

Supplementary Fig.~S2(a) shows the SSIM distributions of the reconstructed images at different mistriggering fractions. We observed a slight but consistent decline in reconstruction quality as the fraction of corrupted k-space lines increased. The representative qualitative results (Fig.~S2(b)-(f)) show reconstructions from a patient with suspected acute myocarditis. The temporal effects of mistriggering were more obvious in the reconstructed y-t plane images. At mistriggering fractions of 0.2 and 0.3, SDUM and CardioMM still preserved the overall anatomy reasonably well in the x-y images, but temporal inconsistencies were clearly visible in the temporal dynamics (y-t plane). RAM reconstructions remained blurred, and temporal inconsistencies were also not eliminated. Since the corrupted k-space lines were selected randomly across k-space, the severity of spatial artifacts in individual x-y images was not necessarily monotonic with the mistriggering fraction, as illustrated in Supplementary Fig.~S2(b).

Fig.~\ref{fig:results_motion_respiratory} shows the effect of simulated respiratory motion on reconstruction performance. For all three models, SSIM noticeably decreased as motion amplitude increased. At $A=10$ pixels, the mean SSIM dropped below 0.4. In Fig.~~\ref{fig:results_motion_respiratory}(b), the fully-sampled images illustrate the simulated respiratory artifacts, which became more severe as displacement increased. We assumed that motion aliasing, which appeared visually similar to undersampling-related artifacts, might be suppressed by the visual encoders or regularizers of the foundation models. However, these motion-induced artifacts remained visible after reconstruction with each investigated model. At $A=5$ and $10$ pixels, SDUM and CardioMM showed error map patterns similar to those present in the fully-sampled images, indicating that the reconstructions remained consistent with the underlying motion-corrupted measurements. RAM remained blurred across all motion levels and degraded further as motion amplitude increased.

\subsection{Signal fidelity sensitivity}
\begin{figure}
    \centering
    \includegraphics[width=\textwidth]{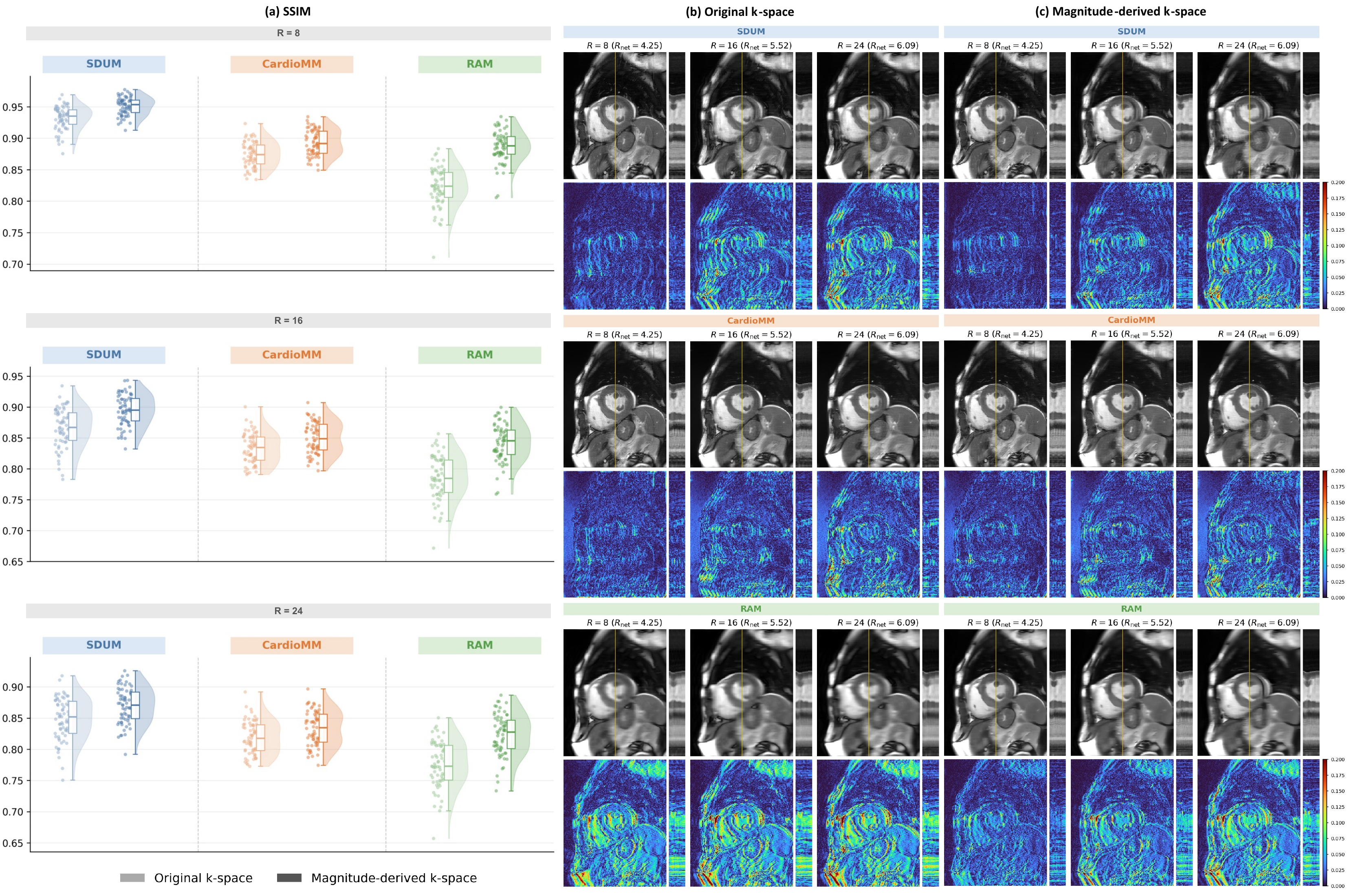}
    \caption{\textbf{Reconstruction performance using original and magnitude-derived k-space.} (a) Raincloud plots showing SSIM distributions for SDUM, CardioMM, and RAM across 65 patients using kt-uniform undersampling at a nominal acceleration factor of $R=8,16,24$, comparing reconstructions from the original k-space and magnitude-derived synthetic k-space. The plots combine individual data points, distribution densities, and boxplots. (b, c) Representative reconstructions and corresponding absolute error maps obtained from the original k-space (b) and magnitude-derived synthetic k-space (c). Effective acceleration factors ($R_{\mathrm{net}}$) are indicated in parentheses above each reconstruction. The yellow lines in the x-y images indicate the corresponding y-t planes.}
    \label{fig:results_magnitude}
\end{figure}

The evaluated models showed limited sensitivity to global signal scaling in the in-house cine MR data (Supplementary Fig.~S3). Across the tested scaling factors, the reconstructed images remained similar for all models. However, a different behavior was observed for RAM in a separate fastMRI brain experiment (Supplementary Fig.~S4). In this experiment, we converted the multi-coil fastMRI brain data to single-coil data using virtual coil combination (VCC), following the processing pipeline used for RAM~\cite{terris2025reconstruct}. Without additional scaling ($\alpha=1$), RAM failed to produce a meaningful reconstruction. However, when we increased the input scale by $\alpha=200$, RAM could produce a valid reconstruction with anatomical details. This dataset-specific scaling sensitivity warrants concern and triggered an in-depth model analysis if a systemic error persists or model-specific internal preprocessing pipelines require adjustments. Further inspection found that the RAM implementation in DeepInverse determines whether to rescale the input using a fixed threshold on the maximum component of the k-space measurement. For the original multi-coil fastMRI brain input, the maximum value was 0.00437, below the internal threshold of 0.025, which triggered an amplification ($\alpha=1/0.00437\approx229$) and increased the p99 value of the corresponding input image from $1.70\times10^{-4}$ to $3.90\times10^{-2}$. However, after virtual coil combination, the single-coil k-space exceeded the threshold and therefore bypassed this rescaling, although its image-domain input remained small (p99 value$=1.06\times10^{-3}$), resulting in failed reconstruction. As the exact VCC implementation used for RAM training was not publicly available, we adopted the VCC implementation from a previous study by Comby et al.~\cite{comby2025robust}. 

Fig.~\ref{fig:results_magnitude} compares reconstruction performance using the original acquired k-space and magnitude-derived k-space at nominal acceleration factors of $R=8,16,24$. Across all three models, magnitude-derived k-space resulted in a better reconstruction with higher SSIM values. SDUM and CardioMM showed relatively modest changes, while RAM showed a pronounced improvement with magnitude-derived k-space. This difference was also evident qualitatively. The reconstructed images of RAM were sharper, particularly at $R=8$, with better preservation of anatomical boundaries and fewer residual artifacts. 

\subsection{Generalization across datasets}
\begin{figure}[p]
	\centering
	\includegraphics[width=\textwidth]{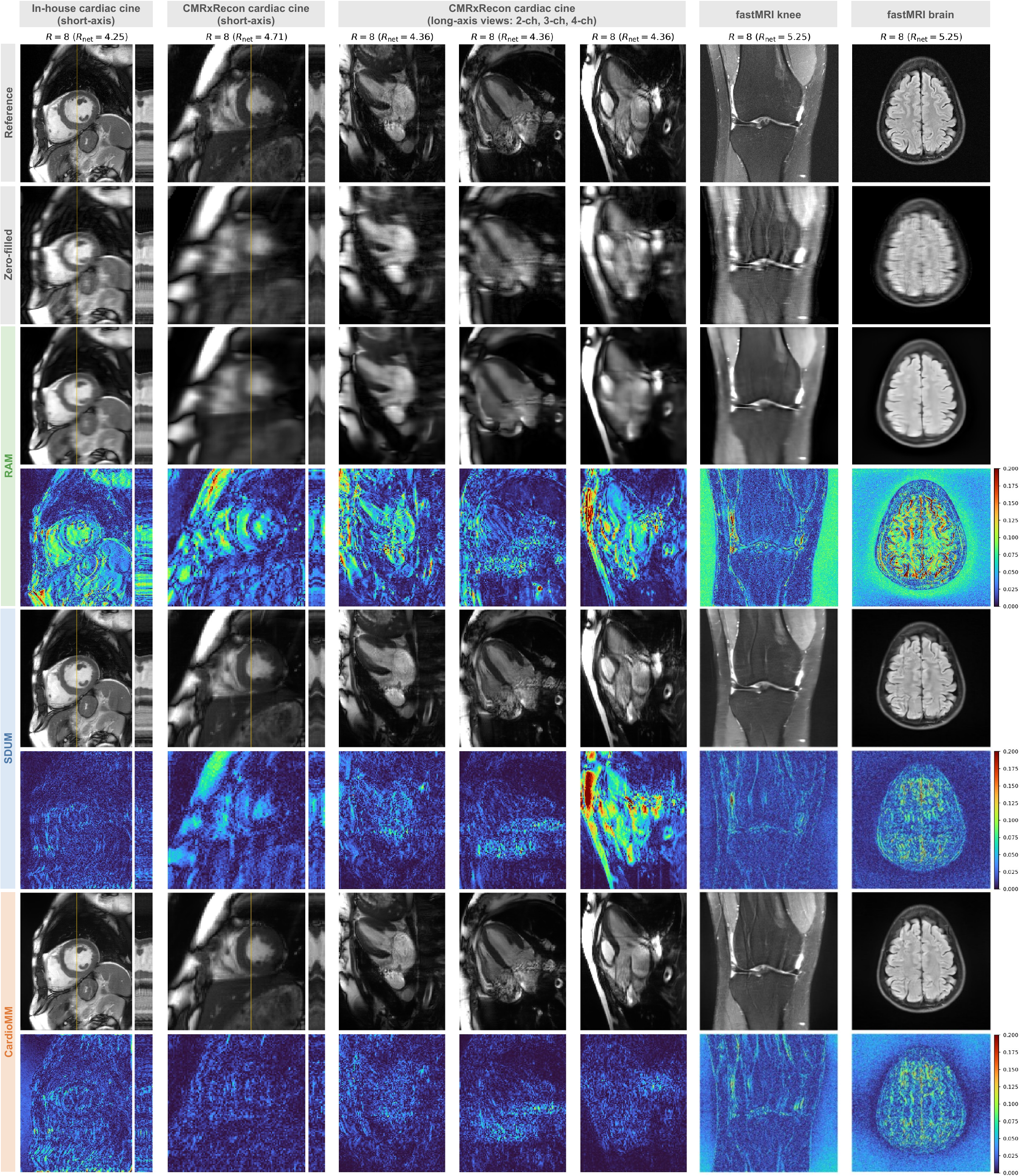}
	\caption{\textbf{Generalization across datasets, anatomies, and cardiac views.} Representative reconstruction results for SDUM, CardioMM, and RAM across in-house cardiac cine MRI, CMRxRecon cardiac cine MRI, fastMRI fat-suppressed PD-weighted TSE knee, and FLAIR TSE brain data. The cardiac cine examples include short-axis imaging from the in-house and CMRxRecon datasets, as well as CMRxRecon long-axis 2-chamber, 3-chamber, and 4-chamber views. Dynamic cardiac data were retrospectively undersampled using kt-uniform sampling, whereas the static fastMRI knee and brain data used Cartesian uniform undersampling, all at a nominal acceleration factor of $R=8$. Effective acceleration factors ($R_{\mathrm{net}}$) are indicated in parentheses. The yellow lines in the x-y images indicate the corresponding y-t planes.}
	\label{FIG:result_dataset}
\end{figure}

Fig.~\ref{FIG:result_dataset} shows representative reconstructions across different datasets, cardiac views, and anatomies. RAM removed some artifacts compared to the zero-filled input images, but its reconstruction remained noticeably blurred across all datasets. Similar blurring was also observed in the brain reconstructions with substantial loss of fine anatomical detail, even though brain MRI was included in the training data. SDUM and CardioMM preserved the overall cardiac anatomy across both the in-house and CMRxRecon cine datasets, including different cardiac views. When applied to out-of-domain anatomies, both models were still able to recover the anatomical structures. However, the reconstructed images contained spurious structures that did not exist in the reference images. Some of these structures appeared to result from enhanced aliasing artifacts.

\afterpage{\clearpage}

\section{Discussion}
In this work, we developed MRBench as a standardized and extensible framework for evaluating pretrained foundation models for MRI reconstruction. Recent reconstruction foundation models have been developed using different datasets, preprocessing strategies, and network architectures, making their reported performance difficult to compare directly. MRBench provides a common evaluation environment in which different datasets and test conditions can be applied across models while preserving the model-specific inference procedures. Together with the public implementation and interactive dashboard, MRBench provides a reproducible basis for characterizing the practical applicability of reconstruction foundation models before broader deployment. The framework is designed to be extensible, allowing additional reconstruction models and datasets to be incorporated through future integrations. 

Using MRBench, we found that the generalization of reconstruction foundation models depends strongly on the test conditions. The three evaluated models showed different operating boundaries when acquisition and signal conditions were varied. Some changes, such as increasing acceleration, led to gradual degradation in reconstruction quality, as also a-priori expected. However, models also turned out to be very sensitive to some other inference settings (as observed for SDUM and CardioMM with VISTA sampling without a fully-sampled central calibration region; or for global scaling in RAM), which in some scenarios also resulted in failed inference. 

Among successful reconstructions, the most accurate models were not always the most robust. SDUM achieved the best reconstruction quality at $R=8$, but showed the largest performance change as acceleration increased (Fig.~\ref{FIG:result_acc}). SDUM was also more sensitive to changes in SNR than RAM, which was more stable across noise levels but yielded lower reconstruction quality  (Supplementary Fig.~S1). Therefore, our benchmark is intended not only to compare reconstruction quality across models, but also to characterize how each model performs as the operating conditions change. To facilitate such comparison, MRBench provides an interactive dashboard with a continuously updated leaderboard, allowing users to select experimental conditions and evaluation metrics of interest and compare model performance accordingly.

The sampling experiments highlight the importance of acquisition compatibility with the reconstruction foundation models. Different sampling patterns resulted in variations in reconstruction quality (Fig.~\ref{FIG:result_sampling}). More importantly, the absence of a fully-sampled ACS region in sampling patterns, such as VISTA sampling, led to failed inference in SDUM and CardioMM (Fig.~\ref{FIG:result_acs}). Both models incorporate coil sensitivity estimation into the reconstruction pipelines, reducing the need for users to provide sensitivity maps separately, but this also imposes limitations on the sampling scheme. Therefore, broader generalization may require future foundation models to support diverse sampling and calibration configurations. For instance, by allowing the model to estimate sensitivity maps internally when calibration data is available and supplying them externally when it is not.

The scaling sensitivity experiments revealed a scale-dependent failure mode in RAM. Although global scaling of k-space had little effect on the in-house cardiac cine data over the tested range (Supplementary Fig.~S3), RAM failed to produce a meaningful reconstruction for the single-coil fastMRI brain data without scaling (Supplementary Fig.~S4). The in-house cine data did not show the same failure over the tested scaling range because the inputs did not fall into the same unfavorable numerical regime, in which the k-space measurement exceeds the rescaling threshold while the corresponding image-domain input remains too small for effective reconstruction. SDUM avoids this scaling failure by applying complex z-score normalization to the input image before reconstruction and recovering the original scale afterward. Similarly, CardioMM performs z-score normalization to each sample within the network and rescales the output back to the original scale after reconstruction. These results show that numerical scale handling is an important part of the inference pipeline. Such pre-processing steps, however, may not be apparent to end-users, particularly when the implementation is not openly accessible. Users may therefore assume that a pretrained foundation model can handle the input data properly, only to encounter unexpected failures when hidden preprocessing assumptions are violated. Future foundation models may therefore benefit from a transparent preprocessing pipeline and from considering image normalization that adapts to each input rather than relying on fixed absolute thresholds.

The models reconstructed magnitude images more accurately from magnitude-derived k-space than from acquired k-space, with the largest gain observed for RAM (Fig.~\ref{fig:results_magnitude}). Whether this gain reflects the preprocessing used during training remains unknown. Publicly documenting how training inputs are generated would help assess this possibility. All three models process the real and imaginary components in two channels using real-valued layers. This representation retains phase information, while the layers can learn how to combine the channels without the constraints of complex-valued linear operations. RAM was trained with a weighted $L_1$ loss on complex-valued images, whereas SDUM and CardioMM used magnitude-domain losses. Our metrics assess magnitude reconstruction quality, the clinical target in cardiac cine MRI. They cannot establish whether removing the acquired image phase caused the observed gains or how accurately the models recover phase from acquired k-space.

The motion experiments further showed that successful reconstruction does not necessarily imply correction of acquisition-induced corruptions. Temporal inconsistency from mistriggering (Supplementary Fig.~S2) and ghosting from respiratory motion (Fig.~\ref{fig:results_motion_respiratory}) remained visible after reconstruction. Although these artifacts were visually similar to undersampling artifacts, the learned image network components did not simply remove them. In other words, the applied data consistency preserved the motion artifacts in the acquired k-space. This behavior also reflects fidelity to structures already encoded in the acquired measurements. A related stability experiment by~\cite{antun2020instabilities} showed that reconstruction networks should be able to preserve structural changes inserted into the input images. A dedicated structural fidelity experiment would require a separate evaluation design and quantitative measures of feature preservation, and is therefore left for future work.


Generalization across anatomies remains a challenge for the evaluated models. SDUM and CardioMM preserved overall cardiac anatomy across the evaluated cine datasets, including in-house data acquired at a different center than the training data and multiple cardiac views from the CMRxRecon dataset. However, when applied to OOD brain and knee anatomies, both models can still reconstruct the anatomical structures but introduce additional structural artifacts (Fig.~\ref{FIG:result_dataset}). It should be noted that both models were developed specifically for cardiac MRI, so brain and knee data represent extrapolations beyond their intended application domain. RAM further showed that anatomical similarity alone does not necessarily yield effective reconstruction. Although brain MRI was included in its training data, reconstruction of the multi-coil fastMRI brain data remained substantially blurred, with loss of fine anatomical detail. Again, a mismatch between training and test data (i.e., the preprocessing pipelines) may explain the poor reconstruction quality, as RAM was trained on virtual coil-combined single-coil fastMRI data, whereas our evaluation used the original multi-coil measurements. Overall, these results suggest that model performance is more reliable when the test data remains close to the development and training domains. In this context, domain similarity depends not only on anatomy but also on other factors, such as input representation and acquisition settings. 

\section{Limitations}
Several limitations of this study should be acknowledged. First, although the evaluated models represent different network designs and development domains, the current version of MRBench includes only three foundation models. Therefore, MRBench was designed to be extensible, and a future version will incorporate additional models, such as the latest version of RAM 2~\cite{blurlabs_ram2}. Additionally, the current study focuses primarily on foundation models. Incorporating task-specific reconstruction models would provide an important reference for determining whether broader generalization comes at the cost of reconstruction performance.

Second, the current study focuses primarily on cardiac cine MRI, but MRBench also supports other MRI datasets. Future extensions will include datasets covering diverse anatomies, imaging contrasts, acquisition protocols, and data sources. These extensions will not only enable broader evaluation of model generalization, but also further establish the applicability and reproducibility of MRBench.

Third, all experiments conducted in this study were performed using retrospectively undersampled data to enable systematic and controlled evaluation across models and acquisition conditions. While this design facilitates reproducible comparisons, prospective evaluation on clinically acquired accelerated data will still be necessary to assess model performance. Furthermore, the current evaluation primarily focuses on image quality metrics. Clinically relevant evaluation, such as ventricular volume and function measurements, should be incorporated to assess diagnostic accuracy. The benchmark will also be extended to evaluate complex-valued signal fidelity for applications where phase information matters, such as flow imaging. The public MRBench dashboard will be updated as new reconstruction models, datasets, and evaluation tasks become available.

\section{Conclusion}
In this study, we introduced MRBench, a standardized and reproducible framework for evaluating the performance, robustness, and generalization of pretrained MRI reconstruction foundation models. Through a set of controlled experimental conditions implemented in MRBench, we found that the evaluated models exhibited distinct operating boundaries. Changes in acquisition conditions, signal fidelity, and data domains could lead to performance variation or even failed reconstruction. MRBench provides a systematic way to identify the strengths, limitations, and deployment constraints of each model. Future extensions will provide additional models, datasets, and evaluations to support a more reliable translation of reconstruction foundation models into clinical workflows.

\section*{Acknowledgments}
The work was supported by the Deutsche Forschungsgemeinschaft (DFG, German Research Foundation) under Germany’s Excellence Strategy – EXC 2064/1 – Project number 390727645, and by the de.NBI Cloud within the German Network for Bioinformatics Infrastructure (de.NBI) and ELIXIR-DE (Forschungszentrum Jülich and W-de.NBI-001, W-de.NBI-004, W-de.NBI-008, W-de.NBI-010, W-de.NBI-013, W-de.NBI-014, W-de.NBI-016, W-de.NBI-022).




\printcredits

\section*{Availability of data and materials}
The in-house cardiac cine MRI data are not publicly available due to patient privacy and ethical restrictions. The CMRxRecon and fastMRI datasets used in this study are publicly available. The MRBench source code is publicly available at https://github.com/lab-midas/MRBench, and the interactive benchmark dashboard is available at https://lab-midas.github.io/MRBench/.

\section*{Declaration of Generative AI and AI-assisted technologies in the writing process}
During the preparation of this work, the authors used ChatGPT and Codex (OpenAI) to assist with language editing, improving the clarity of the manuscript, and code debugging. The scientific analyses, results, interpretations, and conclusions were determined by the authors. The authors reviewed and edited the content as needed and take full responsibility for the content of the published article.

\bibliographystyle{cas-model2-names}

\bibliography{mrbench}



\clearpage
\section*{Supplementary Material}

\vspace{2cm}

\setcounter{figure}{0}
\renewcommand{\thefigure}{S\arabic{figure}}

\begin{center}
    \centering
    \includegraphics[width=\textwidth]{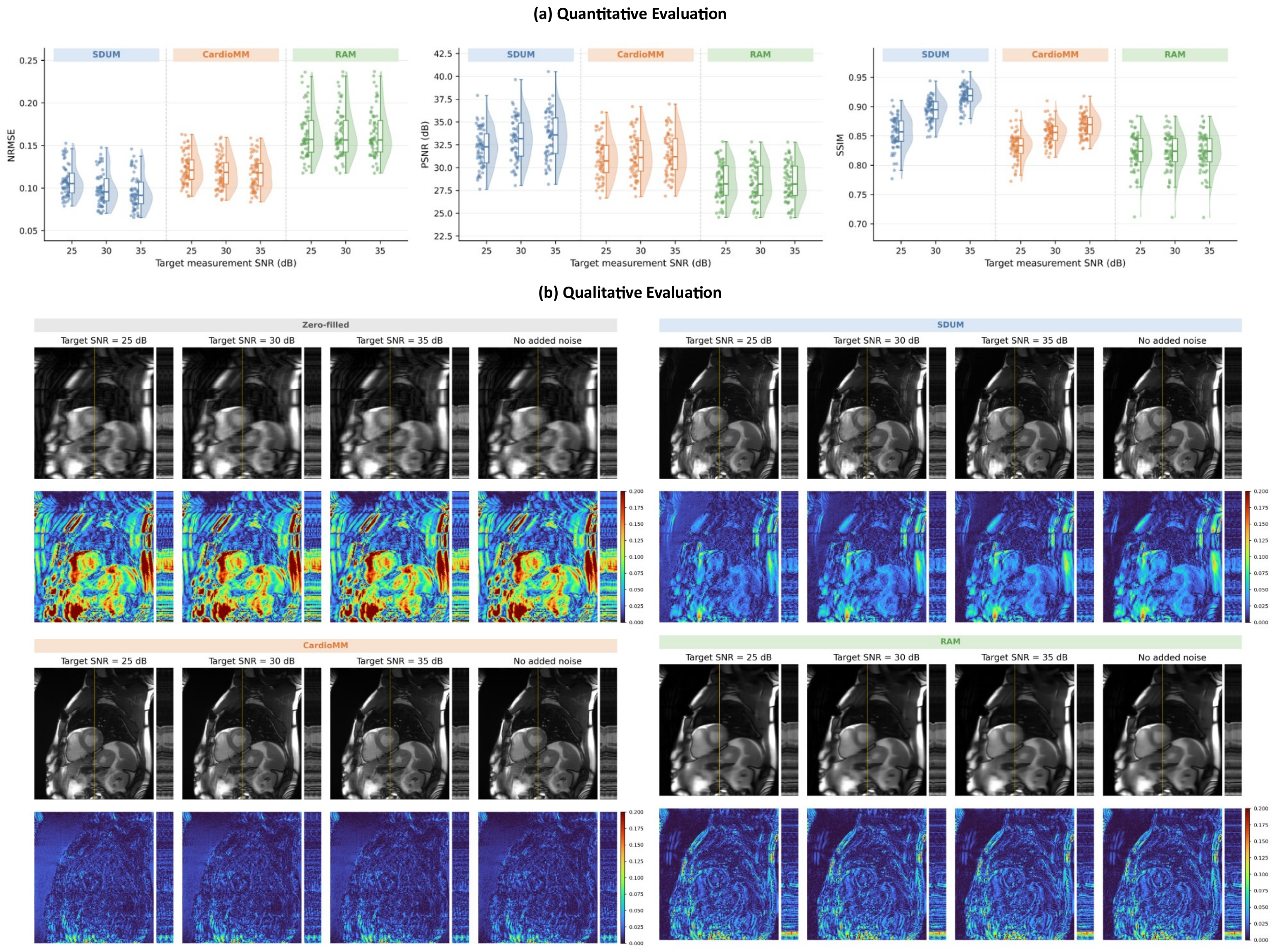}
    \captionof{figure}{\textbf{Reconstruction performance under different measurement noise levels.} (a) Raincloud plots showing NRMSE, PSNR, and SSIM for SDUM, CardioMM, and RAM across 65 patients from an in-house cardiac cine MRI dataset using kt-uniform undersampling at a nominal acceleration factor of $R=8$. Complex Gaussian noise was added to the acquired multi-coil k-space to achieve target measurement SNRs of 25, 30, and 35 dB. The plots combine individual data points, distribution densities, and boxplots. (b) Representative results from one patient at the three target SNR levels and under the corresponding no-added-noise condition, including zero-filled reconstructions and reconstructed images from the three models along with their corresponding absolute error maps. The yellow lines in the x-y images indicate the corresponding y-t planes.}
    \label{fig:results_noise}
\end{center}

\begin{figure}[htbp]
    \centering
    \includegraphics[width=\textwidth]{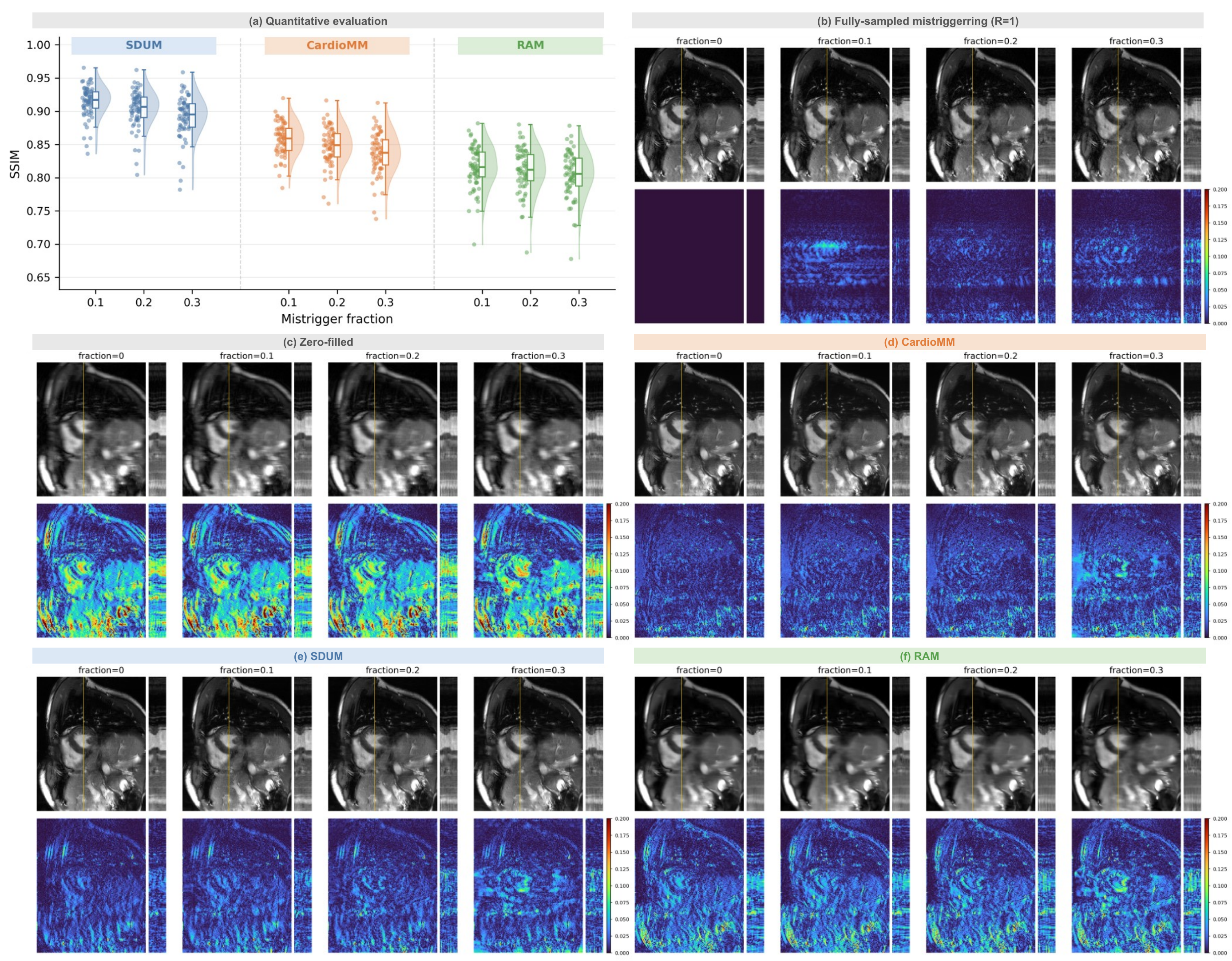}
    \caption{\textbf{Reconstruction performance under simulated mistriggering motion.} (a) Raincloud plots showing SSIM distributions for SDUM, CardioMM, and RAM across 65 patients using kt-uniform undersampling at a nominal acceleration factor of $R=8$, with mistriggering fractions of 0.1, 0.2, and 0.3. The plots combine individual data points, distribution densities, and boxplots. (b) Fully-sampled images with simulated motion, illustrating the motion corruption before undersampling. (c-f) Representative results from one patient, including zero-filled reconstructions and reconstructions from CardioMM, SDUM, and RAM, respectively, along with their corresponding absolute error maps. The mistriggering fraction denotes the proportion of k-space lines corrupted by cardiac phase mismatch. The yellow lines in the x-y images indicate the corresponding y-t planes.}
    \label{fig:results_motion_mistrigger}
\end{figure}

\begin{figure}[htbp]
    \centering
    \includegraphics[width=\textwidth]{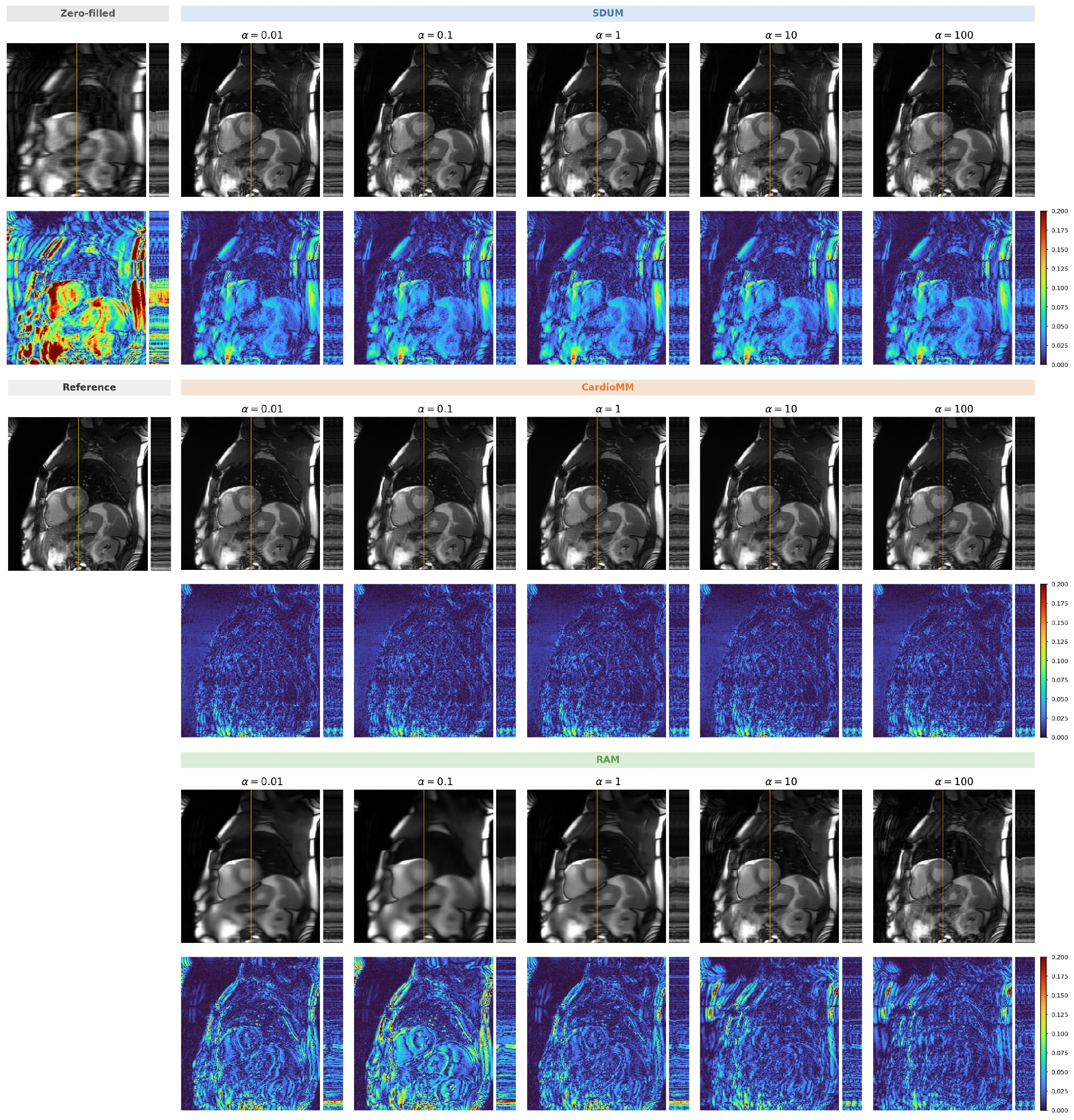}
    \caption{\textbf{Reconstruction performance under different global scaling factors.} Representative results from one patient using kt-uniform undersampling at a nominal acceleration factor of $R=8$. Global scaling factors of $\alpha=0.01,0.1,1,10,$ and $100$ were applied to the input k-space before reconstruction with SDUM, CardioMM, and RAM. The fully-sampled reference and zero-filled reconstruction are shown along with the reconstructed images from each model and their corresponding absolute error maps. The yellow lines in the x-y images indicate the corresponding y-t planes.}
    \label{fig:results_scaling}
\end{figure}

\begin{figure}[htbp]
    \centering
    \includegraphics[width=0.7\textwidth]{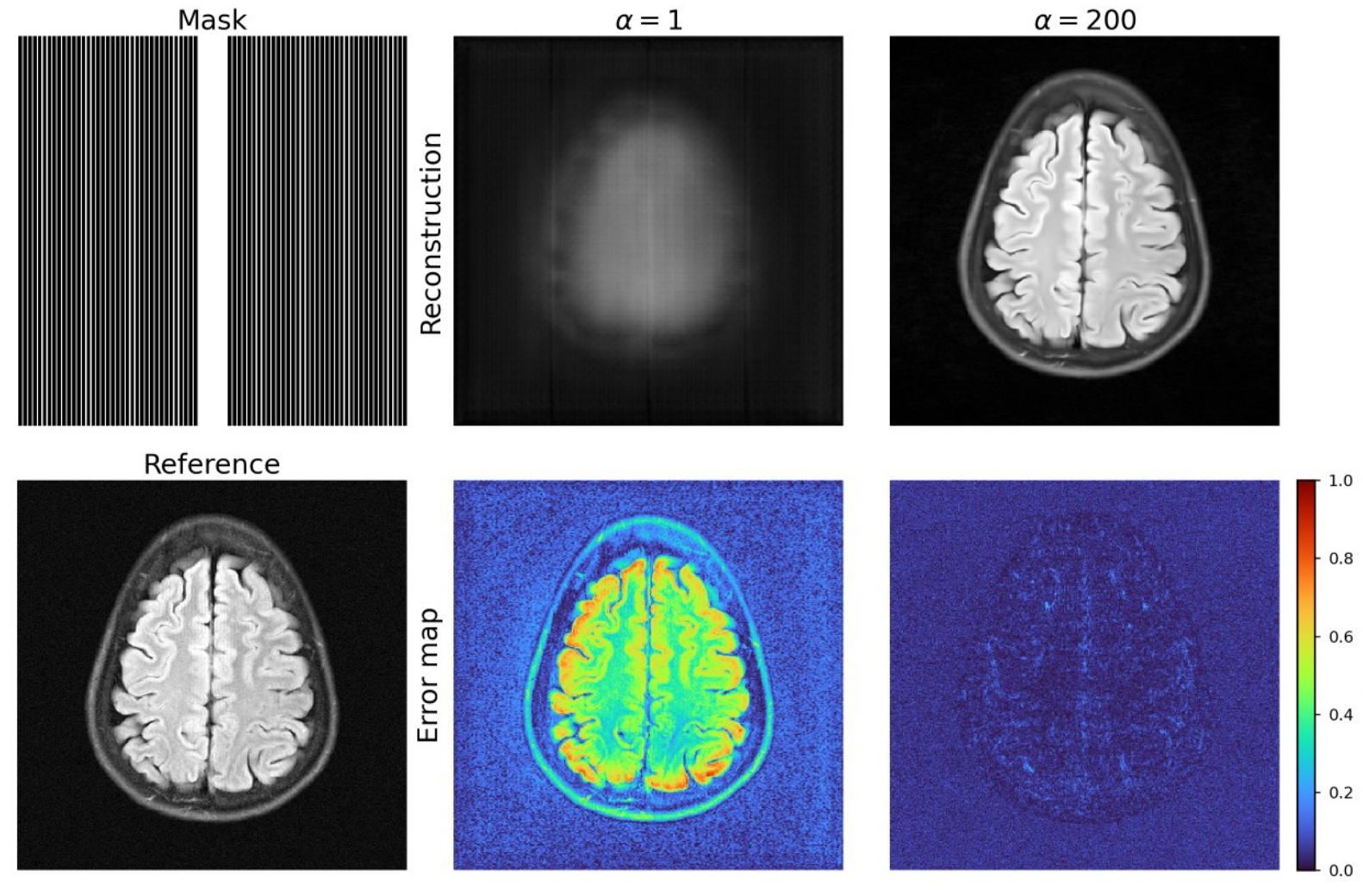}
    \caption{\textbf{Sensitivity of RAM to global input scaling in fastMRI brain data.} A representative multi-coil fastMRI brain slice was converted to a complex single-coil image using virtual coil combination and retrospectively undersampled with a Cartesian uniform mask at a nominal acceleration factor of $R=4$ ($R_{\mathrm{net}}=3.27$ with $n_{\mathrm{calib}}=24$). Reconstructions are shown for global scaling factors of $\alpha=1$ and $\alpha=200$, along with the corresponding absolute error maps.}
    \label{fig:results_fastMRI_ram_scaling}
\end{figure}

\end{document}